\documentclass[runningheads]{llncs}
\usepackage{amssymb}
\usepackage{amsmath}
\usepackage{graphicx}
\usepackage{url}
\usepackage[export]{adjustbox}
\usepackage{array}
\usepackage[labelformat=empty]{subfig}
\usepackage{color,soul} 
\soulregister\ref{7} 
\soulregister\cite{7} 
\usepackage{float} 
\usepackage[colorlinks]{hyperref}
\usepackage[colorinlistoftodos]{todonotes}
\usepackage{booktabs} 
\usepackage{rotating}
\usepackage{multirow}
\usepackage{transparent}
\usepackage{tablefootnote}
\usepackage{algorithm}
\usepackage{algpseudocode}
\usepackage{subfig}

\usepackage{bm} 
\usepackage{colortbl}
\usepackage{tabulary}

\usepackage{booktabs, multirow} 
\usepackage{soul}
\usepackage{changepage,threeparttable} 

\begin{document}
\mainmatter              

\title{Federated Learning for Breast Density Classification: A Real-World Implementation}
\titlerunning{Federated Learning for Breast Density Classification}  %

\author{Holger R. Roth\inst{1}, 
Ken Chang\inst{2}, 
Praveer Singh\inst{2}, 
Nir Neumark\inst{2}, 
Wenqi Li\inst{1}, 
Vikash Gupta\inst{3}, 
Sharut Gupta\inst{2}, 
Liangqiong Qu\inst{4},  
Alvin Ihsani\inst{1}, 
Bernardo C. Bizzo\inst{2}, 
Yuhong Wen\inst{1}, 
Varun Buch\inst{2}, 
Meesam Shah\inst{5}, 
Felipe Kitamura\inst{6}, 
Matheus Mendon\c{c}a\inst{6}, 
Vitor Lavor\inst{6}, 
Ahmed Harouni\inst{1},  
Colin Compas\inst{1}, 
Jesse Tetreault\inst{1}, 
Prerna Dogra\inst{1}, 
Yan Cheng\inst{1}, 
Selnur Erdal\inst{3}, 
Richard White\inst{3}, 
Behrooz Hashemian\inst{2}, 
Thomas Schultz\inst{2}, 
Miao Zhang\inst{4},  
Adam McCarthy\inst{2}, 
B. Min Yun\inst{2}, 
Elshaimaa Sharaf\inst{2}, 
Katharina V. Hoebel\inst{7}, 
Jay B. Patel\inst{7}, 
Bryan Chen\inst{7}, 
Sean Ko\inst{7}, 
Evan Leibovitz\inst{2}, 
Etta D. Pisano\inst{2}, 
Laura Coombs\inst{5}, 
Daguang Xu\inst{1}, 
Keith J. Dreyer\inst{2}, 
Ittai Dayan\inst{2}, 
Ram C. Naidu\inst{2}, 
Mona Flores\inst{1}, 
Daniel Rubin\inst{4}, 
Jayashree Kalpathy-Cramer\inst{2}} 

\institute{
NVIDIA, Santa Clara, USA \and
Massachusetts General Hospital, Boston, USA \and
Mayo Clinic, Jacksonville, USA \and
Stanford University, Stanford, USA \and
American College of Radiology, Reston, USA \and
Diagn{\'o}sticos da Am{\'e}rica (DASA), S{\~a}o Paulo, Brazil \and
Massachusetts Institute of Technology, Cambridge, USA
}
\authorrunning{Holger R. Roth et al.}

\maketitle              

\begin{abstract}
Building robust deep learning-based models requires large quantities of diverse training data. In this study, we investigate the use of federated learning (FL) to build medical imaging classification models in a real-world collaborative setting. Seven clinical institutions from across the world joined this FL effort to train a model for breast density classification based on Breast Imaging, Reporting \& Data System (BI-RADS). We show that despite substantial differences among the datasets from all sites (mammography system, class distribution, and data set size) and without centralizing data, we can successfully train AI models in federation. The results show that models trained using FL perform 6.3\% on average better than their counterparts trained on an institute's local data alone. Furthermore, we show a 45.8\% relative improvement in the models' generalizability when evaluated on the other participating sites' testing data.
\end{abstract}


\section{Introduction}
\label{sec:intro}
Advancements in medical image analysis over the last several years have been dominated by deep learning (DL) approaches. However, it is well known that DL requires large quantities of data to train robust and clinically useful models \cite{dunnmon2019assessment,chang2020multi}. Often, hospitals and other medical institutes need to collaborate and host centralized databases for the development of clinically useful models. This overhead can quickly become a logistical challenge and usually requires a time-consuming approval process due to data privacy and ethical concerns associated with data sharing in healthcare \cite{larson2020ethics}. Even when these challenges can be addressed, data is valuable, and institutions may prefer not to share full datasets. Furthermore, medical data can be large, and it may be prohibitively expensive to acquire storage for central hosting \cite{chang2018distributed}. One approach to combat the data sharing hurdles is federated learning (FL) \cite{mcmahan2017communication}, where only model weights are shared between participating institutions without sharing the raw data.

To investigate the performance of FL in the real world, we conducted a study to develop a breast density classification model using mammography data. An international group of hospitals and medical imaging centers joined this collaborative effort to train the model in purely data-decentralized fashion without needing to share any data. This is in contrast to previous studies in which the FL environment was only simulated \cite{sheller2018multi,li2019privacy}.
We do not have centralized training experiments as references before starting the FL tasks, which places higher requirements on the robustness of the algorithms and selection of hyper-parameters.
\subsection{Related Works}
\label{related_works}
\paragraph{\textbf{Breast density scoring:}}
The classification of breast density is quintessential for breast imaging to estimate the extent of fibroglandular tissue related to the patient's risk of developing breast cancer \cite{boyd1995quantitative,razzaghi2012mammographic}. Women with a high mammographic breast density ($>$75\%) have a four- to five-fold increase in risk for breast cancer compared to women having a lower breast density \cite{boyd2007mammographic,yaghjyan2011mammographic}. This condition affects roughly half of American women between the ages of 40 to 74 \cite{ho2014dense,sprague2014prevalence}. Patients identified with dense breast tissue may have masked tumors and benefit from supplemental imaging such as MRI or ultrasound \cite{lehman2019mammographic}. High mammographic breast density impairs the sensitivity and specificity of breast cancer screening, possibly because (small) malignant lesions are not detectable even when they are present \cite{ooms2007mammography}. The standard evaluation metric for reporting breast density is the Breast Imaging Reporting and Data System (BI-RADS), based on 2D mammography \cite{sickles2013acr}. Scans are categorized into one of four classes: (a) fatty, (b) scattered, (c) heterogeneously dense, and (d) extremely dense.

Due to the subjective nature of the BI-RADS criteria, there can be substantial inter-rater variability between pairs of clinicians. Sprague et al. \cite{sprague2016variation} found that the likelihood of a mammogram being read as dense varies from radiologist to radiologist between 6.3\% to 84.5\%. Ooms et al. find that the overall agreement between four observers (inter-rater agreement) in terms of the overall weighted kappa was 0.77 \cite{ooms2007mammography}. Another study reported the inter-rater variability to be simple kappa = 0.58 among 34 community radiologists \cite{spayne2012reproducibility}. Even the intra-rater agreement in the assessment of BI-RADS breast density can be relatively low. Spayne et al. \cite{spayne2012reproducibility} showed that the intra-rater agreement was below 80\% when evaluating the same mammography exam within a 3‐ to 24‐month period. Recent work on applying DL for mammography breast density classification \cite{lehman2019mammographic} achieved a linear kappa of 0.67 when comparing the DL model's predictions to the assessments of the original interpreting radiologist.
\paragraph{\textbf{Federated Learning:}}
Federated learning has recently been described as being instrumental for the future of digital health \cite{rieke2020future}.
FL enables collaborative and decentralized DL training without sharing any raw patient data \cite{mcmahan2017communication}. Each client in FL trains locally on their data and then submits their model parameters to a server that accumulates and aggregates the model updates from each client. Once a certain number of clients have submitted their updates, the aggregated model parameters are redistributed to the clients, and a new round of local training starts. While out of the scope of this work, FL can also be combined with additional privacy-preserving measures to avoid potential reconstruction of training data through model inversion if the model parameters would be exposed to an adversary \cite{li2019privacy}. Recent works have shown the applicability of FL to medical imaging tasks \cite{sheller2018multi,li2019privacy}. The security and privacy-preserving aspects of federated machine learning in medical imaging have been discussed in more detail by Kaissis et al. \cite{kaissis2020secure}.
\begin{figure*}[htbp]
  \newcommand{\figheight}{1.2cm} 
	\centering
	\begin{tabular}{ccccc}
		\subfloat{\adjincludegraphics[valign=c,height=\figheight]{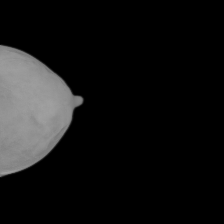}} &
		\hfill
		\subfloat{\adjincludegraphics[valign=c,height=\figheight]{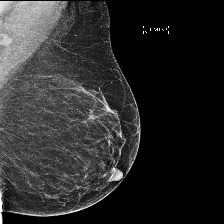}} & 
		\hfill
		\subfloat{\adjincludegraphics[valign=c,height=\figheight]{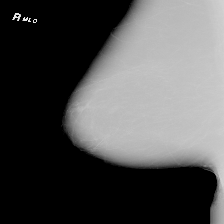}} &
		\hfill
		\subfloat{\adjincludegraphics[valign=c,height=\figheight]{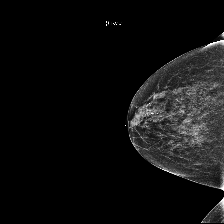}} &
		\hfill
		\subfloat{\adjincludegraphics[valign=c,height=\figheight]{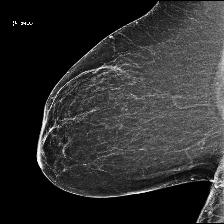}} \\
	\end{tabular}
	\caption{Mammography data examples from different sites after resizing the original images to a resolution of 224 $\times$ 224. No special normalization was applied in order to keep the scanners' original intensity distribution that can be observed in \ref{fig:intensity_distribution}.
	    \label{fig:mammo}}
\end{figure*}
\begin{figure}[h]
 \centering
    \includegraphics[width=0.6\textwidth]{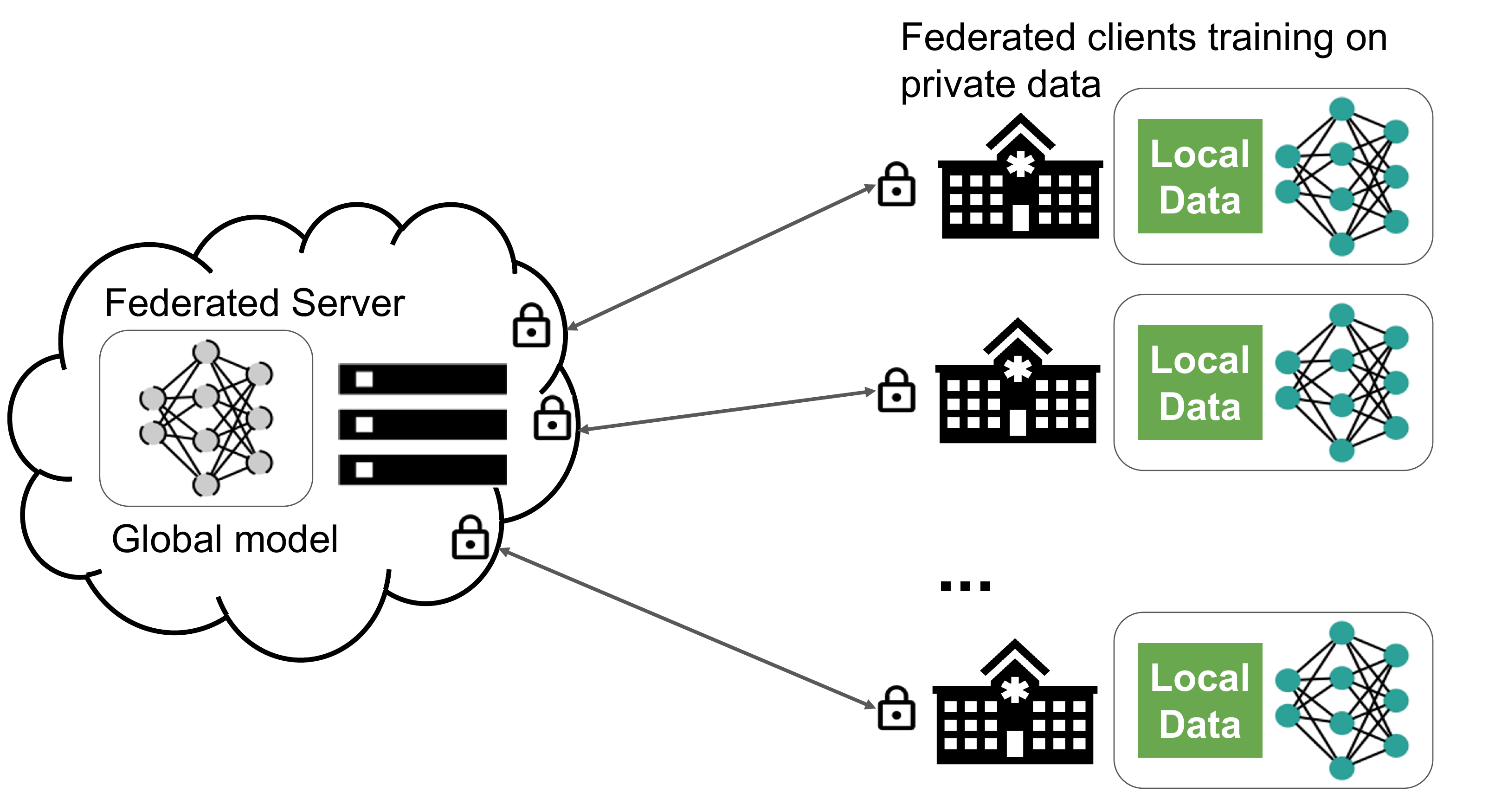}
\caption{
    Federated learning in medical imaging. 
    The central server communicates with clients from multi-national institutions without exchanging any sensitive raw data. Still, the global model benefits from weights and gradients from clients' local models to achieve higher overall performance.
}
\label{fig:fl}
\end{figure}

\section{Method}
\label{sec:method}
\label{sec:experiments}
We implemented our FL approach in a real-world setting with participation from seven international clients. 

\paragraph{\textbf{Datasets:}}
The mammography data was retrospectively selected after Institutional Review Board (IRB) approval as part of standard mammography screening protocols. The BI-RADS breast density class from the original interpreting radiologist was collected from the reports available in the participating hospitals' medical records and includes images from digital screening mammography (Fig. \ref{fig:mammo}). Clients 1 to 3 utilized the multi-institutional dataset previously described in \cite{pisano2005diagnostic}, which was split by the digital mammography system used to acquire the image to account for different dataset sources.

\begin{figure}[htbp]
 \centering
    \includegraphics[width=0.8\textwidth]{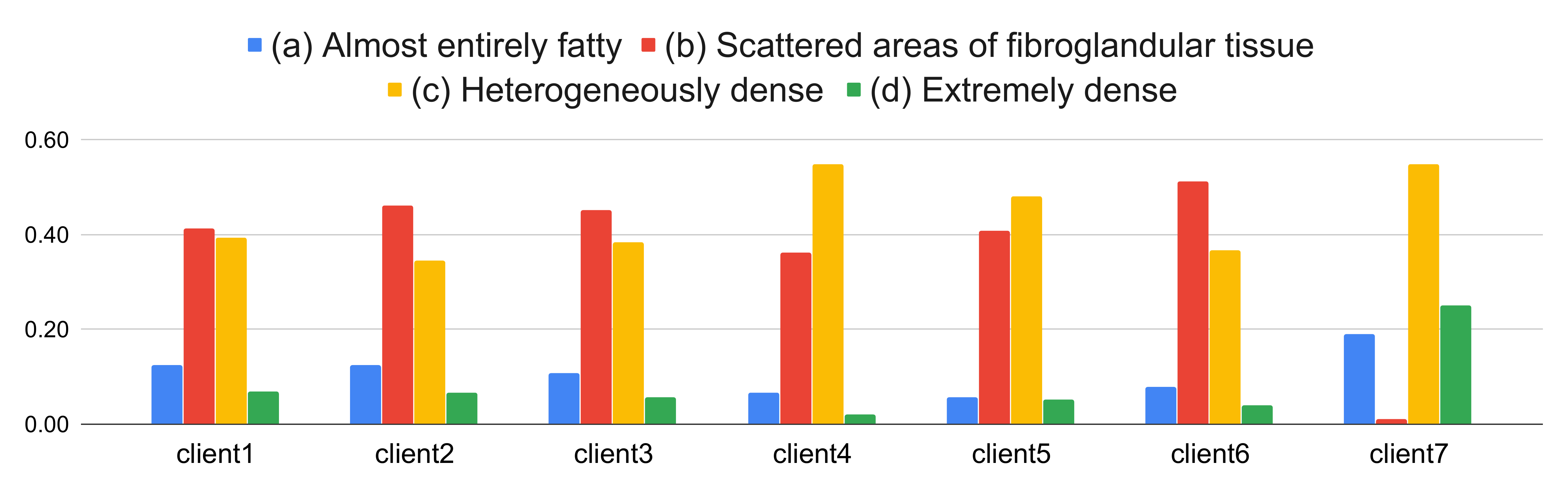}
    \caption{Class distribution at different client sites as a fraction of their total data.}
    \label{fig:mammography_class_distribution}
\end{figure}

Each client's data exhibited their own characteristics of detector type, image resolution, and mammography type. Furthermore, the number of training images varied significantly among clients, as shown in Table \ref{tab:data_characteristics}. The distributions of the different BI-RADS categories were markedly different at some clients but generally followed the distribution known from the literature, with more images in the categories b and c \cite{pisano2005diagnostic}, see Fig. \ref{fig:mammography_class_distribution}. Given these differences that are quite typical for real-world multi-institutional datasets, we can see that the data used in this study is non-independent and identically distributed (non-IID).

\begin{figure}[htbp]
 \centering
    \includegraphics[width=0.8\textwidth]{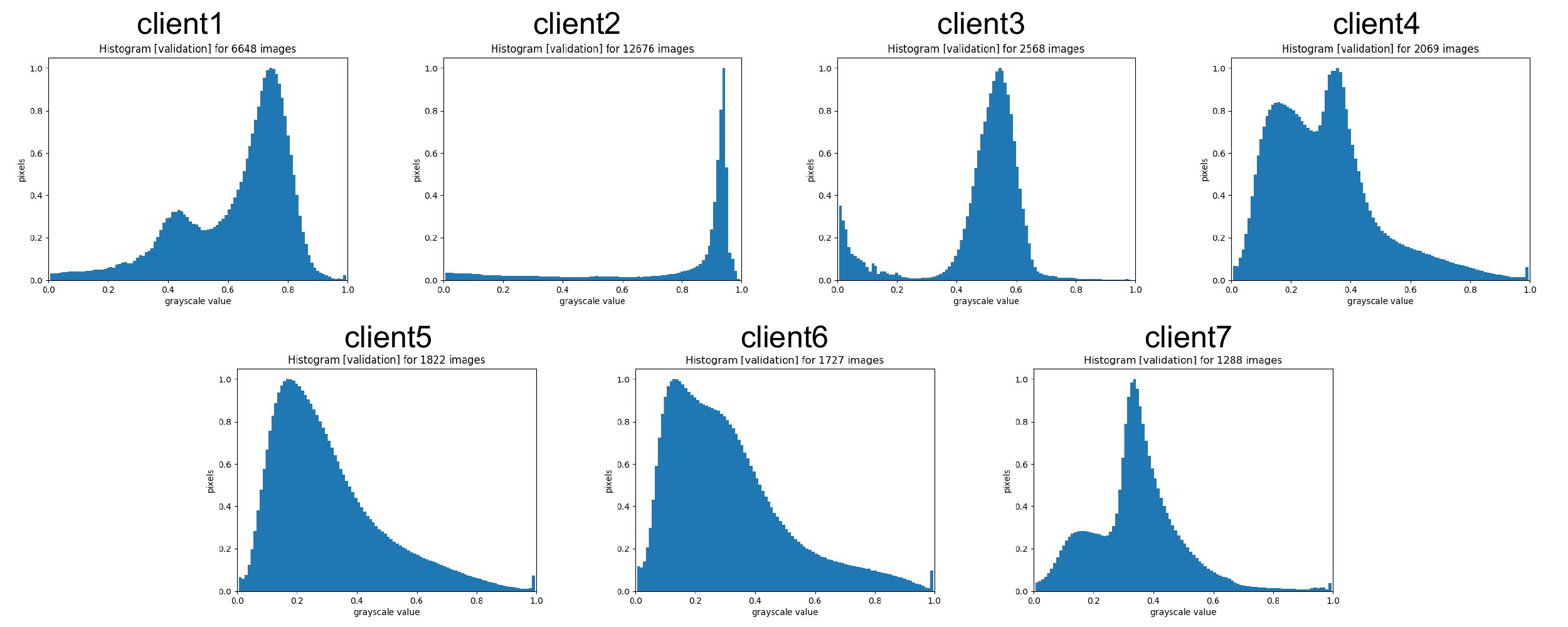} 
    \caption{Intensity distribution at different sites.}%
    \label{fig:intensity_distribution}
\end{figure}

Intensity distributions among different sites also varied markedly, as can be observed in Fig. \ref{fig:intensity_distribution}. This variance is due to the differences in imaging protocols and digital mammography systems used at each data contributing site. No attempt to consolidate these differences was made in our study to investigate the domain shift challenges proposed by this non-IID data distribution.
\begin{table}[]
    \caption{Dataset characteristics at each client. Image resolution is shown in megapixels (MP).}
    \centering
    \scriptsize
    \begin{tabular}{c|c|c|c|c|r|r|r}
        Institution &Image resolution	&Detector type	&Image type	&Bits	&\# Train &\# Val. &\# Test \\
        \hline
        client1 &23.04		&Direct	&2D	&12	&22933 &3366 &6534 \\
        client2 &.02 to 4.39		&Direct	&2D	&12	&8365 &1216 &2568  \\
        client3 &4.39 to 13.63		&Direct	&2D	&14	&44115 &6336 &12676  \\
        client4 &4 to 28 	&Direct/Scintillator	&2D	&12	&7219 &1030 &2069  \\
        client5 &8.48 to 13.63 	&Direct	&2D	&12	&6023 &983 &1822  \\        
        client6 &~8.6 to 13.63	&Direct	&2D	&12	&6874 &853 &1727  \\
        client7 &1 to 136 	&Direct/Scintillator	&2D/tomosynthesis	&10/12	&4021 &664 &1288  \\
    \end{tabular}
    \label{tab:data_characteristics}
\end{table}

\paragraph{\textbf{Client-Server-Based Federated Learning:}}
In its typical form, FL utilizes a client-server setup (Fig. \ref{fig:fl}). Each client trains the same model architecture locally on their data. Once a certain number of clients finishes a round of local training, the updated model weights (or their gradients) are sent to the server for aggregation. After aggregation, the new weights on the server are re-distributed to the clients, and the next round of local model training begins. After a certain number of FL rounds, the models at each client converge. Each client is allowed to select their locally best model by monitoring a certain performance metric on a local hold-out validation set. The client can select either the global model returning from the server after averaging or any intermediate model considered best during local training based on their validation metric. In our experiments, we implement the \texttt{FederatedAveraging} algorithm proposed in \cite{mcmahan2017communication}. While there exist variations of this algorithm to address particular learning tasks, in its most general form, FL tries to minimize a global loss function $\mathcal{L}$ which can be a weighted combination of $K$ local losses $\{\mathcal{L}_k\}_{k=1}^{K}$, each of which is computed on a client $k$'s local data.
Hence, FL can be formulated as the task of finding the model parameters $\phi$ that minimize $L$ given some local data $X_{k} \in X$, where $X$ would be the combination of all local datasets.
\begin{align}
\min_{\phi}\mathcal{L}(X; \phi) \quad \text{with} \quad  \mathcal{L}(X; \phi)=\sum_{k=1}^{K}w_{k}\;\mathcal{L}_{k}(X_{k}; \phi),
\label{eq:formalism}
\end{align}
where $w_k>0$ denotes the weight coefficients for each client $k$, respectively. Note that the local data $X_{k}$ is never shared among the different clients. Only the model weight differences are accumulated and aggregated on the server as shown in Algorithm~\ref{alg:fl}.
\begin{algorithm}[htb]
\caption{Client-server federated learning with \texttt{FederatedAveraging}~\cite{mcmahan2017communication,li2019privacy}. $T$ is the number of federated learning rounds and $n_k$ is the number of \texttt{LocalTraining} iterations minimizing the local loss $\mathcal{L}_{k}(X_{k}; \phi^{(t-1)})$ for a client $k$.}
\label{alg:fl}
\scriptsize
  \begin{algorithmic}[1]
    \Procedure{Federated Learning}{}
      \State{Initialize weights: $\phi^{(0)}$}
      \For{$t \gets 1\cdots T$}
        \For{$client\ k \gets 1\cdots K$}  \Comment{ \textit{Executed in parallel}}
        \State{Send $\phi^{(t-1)}$ to client $k$}
        \State{Receive $(\Delta \phi_k^{(t)}, n_k)$ from client's $\texttt{LocalTraining}(\phi^{(t-1)})$}
        \EndFor
        \State{$\phi_k^{(t)}\gets \phi^{(t-1)} + \Delta \phi_k^{(t)}$}
        \State{$\phi^{(t)}\gets \frac{1}{\sum_k{n_k}}\sum_k{(n_k\cdot \phi_k^{(t)})}$}
      \EndFor
      \State \Return $\phi^{(t)}$
    \EndProcedure
  \end{algorithmic}
\end{algorithm}

In this work, we choose a softmax cross-entropy loss which is commonly used for multi-class classification tasks:
$\mathcal{L}_0 = -\sum_{i=1}^{C} y_i \log{\left(p_i\right)}$;
with $C=4$ being the number of classes. Here, $p_i$ is the predicted probability for a class $i$ from the final softmax activated output layer of our neural network $f(x)$ and $y$ is the one-hot encoded ground truth label for a given image.
%
\paragraph{\textbf{Classification Model \& Implementation:}}
In this work, we do not focus on developing a new model architecture but instead focus on showing how FL works in a real-world collaborative training situation. We implement a DenseNet-121 \cite{huang2017densely} model as a backbone and append a fully-connected layer with four outputs to its last feature layer to classify a mammography image as one of the four BI-RADS categories. The FL framework is implemented in Tensorflow\footnote{\url{https://www.tensorflow.org/}} and utilizes the NVIDIA Clara Train SDK\footnote{\url{https://developer.nvidia.com/clara}} to enable the communication between server and clients as well as to standardize the training configuration among clients. Each client employed an NVIDIA GPU with at least 12 GB memory.

All mammography images were normalized to an intensity range of $[0\dots1]$ and resampled to a resolution of 224 $\times$ 224. We include both left and right breast images and all available views (craniocaudal and mediolateral oblique) in training. Each client separated their dataset into training, validation, and testing sets on the patient level (see Table \ref{tab:data_characteristics}). At inference time, predictions from all images from a given patient were averaged together to give a patient-level prediction.

Each client trained for one epoch before sending their updated model weights to the server for aggregation, and the server waited for all clients before performing a weighted sum of the clients' weight differences. We used initial learning rates of 1e-4 with step-based learning rate decay, Adam optimization for each client, and model weight decay. A mini-batch of size 32 was sampled from the dataset such that all categories were equally represented during training. Random spatial flips, rotations between $\pm$ 45 degrees, and intensity shifts were used as on-the-fly image augmentation to avoid overfitting to the training data. The FL training was run for 300 rounds of local training and weight aggregations, which took about 36 hours. After the FL training is finished, each client's best local model is shared with all other clients and tested on their test data to evaluate the models' generalizability.

In an additional experiment, we use the locally best models each client receives after FL to execute a second round of local fine-tuning based on this model. This additional ``adaptation'' step can improve a client's model on their local data.

\paragraph{\textbf{Evaluation Metric:}}
We utilize Cohen's linear weighed kappa\footnote{\url{https://scikit-learn.org/stable/modules/generated/sklearn.metrics.cohen_kappa_score.html}} to evaluate the locally best models' performance before and after federated learning in comparison with the radiologists' ground truth assessments. The kappa score is a number between -1 and 1. Scores above 0.8 are generally considered very good agreement, while zero or lower would mean no agreement (practically random assignment of labels). A kappa of 0.21 to 0.40, 0.41 to 0.60, and 0.61 to 0.80 represents fair, moderate, and substantial agreement, respectively \cite{landis1977measurement}. The kappa measure has been chosen to be directly comparable to previous literature on breast density classification in mammography \cite{lehman2019mammographic,ooms2007mammography,spayne2012reproducibility}.

\section{Results}

In Table \ref{tab:mammography_generalizibility}, we show the performance of locally best models (selected by best validation score on local data) using local training data alone as well as after federated learning. On average, a 6.3\% relative improvement can be observed when the model is applied to a client's test data (diag. mean). We also observe a general improvement of these best local models applied to the different clients' test data. Here, the generalizability (off-diag. mean) of the models improved by 45.8\% on average.


\begin{adjustwidth}{-0.5 cm}{-0.5 cm}\centering\begin{threeparttable}[!htb]
\caption{Performance of locally best models (selected by best validation score on local data) using (a) local training data alone and (b) after federated learning.}\label{tab:mammography_generalizibility}
\scriptsize
\begin{tabular}{lrrrrrrrrrrrrrrrrrrr}\toprule
&\textbf{(a) Local:} &\multicolumn{7}{c}{\textbf{Test}} & & &\textbf{(b) Federated:} &\multicolumn{7}{c}{\textbf{Test}} \\\cmidrule{2-9}\cmidrule{12-19}
\multirow{8}{*}{\rotatebox[origin=c]{90}{\textbf{Train}}} &\textbf{client} &1 &2 &3 &4 &5 &6 &7 & &\multirow{8}{*}{\rotatebox[origin=c]{90}{\textbf{Train}}} &\textbf{client} &1 &2 &3 &4 &5 &6 &7 \\\midrule
&1 &\cellcolor[HTML]{A8A8A8}{0.62} &0.59 &0.44 &0.02 &0.02 &-0.01 &0.04 & & &1 &\cellcolor[HTML]{A8A8A8}\textbf{0.62} &0.62 &0.48 &0.15 &0.23 &0.24 &0.11 \\
&2 &0.15 &\cellcolor[HTML]{A8A8A8}{0.56} &0.02 &-0.01 &-0.00 &0.00 &-0.01 & & &2 &0.22 &\cellcolor[HTML]{A8A8A8}\textbf{0.65} &0.11 &0.04 &0.00 &0.00 &-0.01 \\
&3 &0.19 &0.01 &\cellcolor[HTML]{A8A8A8}{0.64} &0.02 &0.07 &0.00 &0.05 & & &3 &0.41 &0.17 &\cellcolor[HTML]{A8A8A8}\textbf{0.63} &0.07 &-0.00 &0.01 &-0.01 \\
&4 &0.11 &0.02 &-0.00 &\cellcolor[HTML]{A8A8A8}{0.63} &0.52 &0.61 &0.50 & & &4 &0.06 &0.48 &-0.02 &\cellcolor[HTML]{A8A8A8}\textbf{0.69} &0.57 &0.65 &0.52 \\
&5 &-0.00 &-0.01 &-0.03 &0.54 &\cellcolor[HTML]{A8A8A8}{0.62} &0.65 &0.31 & & &5 &0.24 &0.13 &0.02 &0.64 &\cellcolor[HTML]{A8A8A8}\textbf{0.62} &0.69 &0.52 \\
&6 &0.01 &0.11 &-0.02 &0.49 &0.59 &\cellcolor[HTML]{A8A8A8}{0.71} &0.32 & & &6 &0.23 &0.01 &-0.00 &0.53 &0.68 &\cellcolor[HTML]{A8A8A8}\textbf{0.76} &0.31 \\
&7 &0.03 &0.05 &-0.05 &0.40 &0.37 &0.46 &\cellcolor[HTML]{A8A8A8}{0.69} & & &7 &0.10 &0.21 &0.13 &0.55 &0.44 &0.52 &\cellcolor[HTML]{A8A8A8}\textbf{0.77} \\
\hline
& & & & & & & & & & &\textbf{Global} &0.51 &0.52 &0.49 &0.31 &0.4852 &0.31 &0.0893 \\
\bottomrule
&\multicolumn{7}{c}{\textbf{diag. mean}} &\textbf{0.64} & & &\multicolumn{7}{c}{\textbf{diag. mean}} &\textbf{0.68} \\
&\multicolumn{7}{c}{\textbf{off-diag. mean}} &\textbf{0.18} & & &\multicolumn{7}{c}{\textbf{off-diag. mean}} &\textbf{0.26} \\
\bottomrule
\end{tabular}
\end{threeparttable}\end{adjustwidth}

\begin{figure}[htbp]
 \centering
    \includegraphics[width=0.8\textwidth]{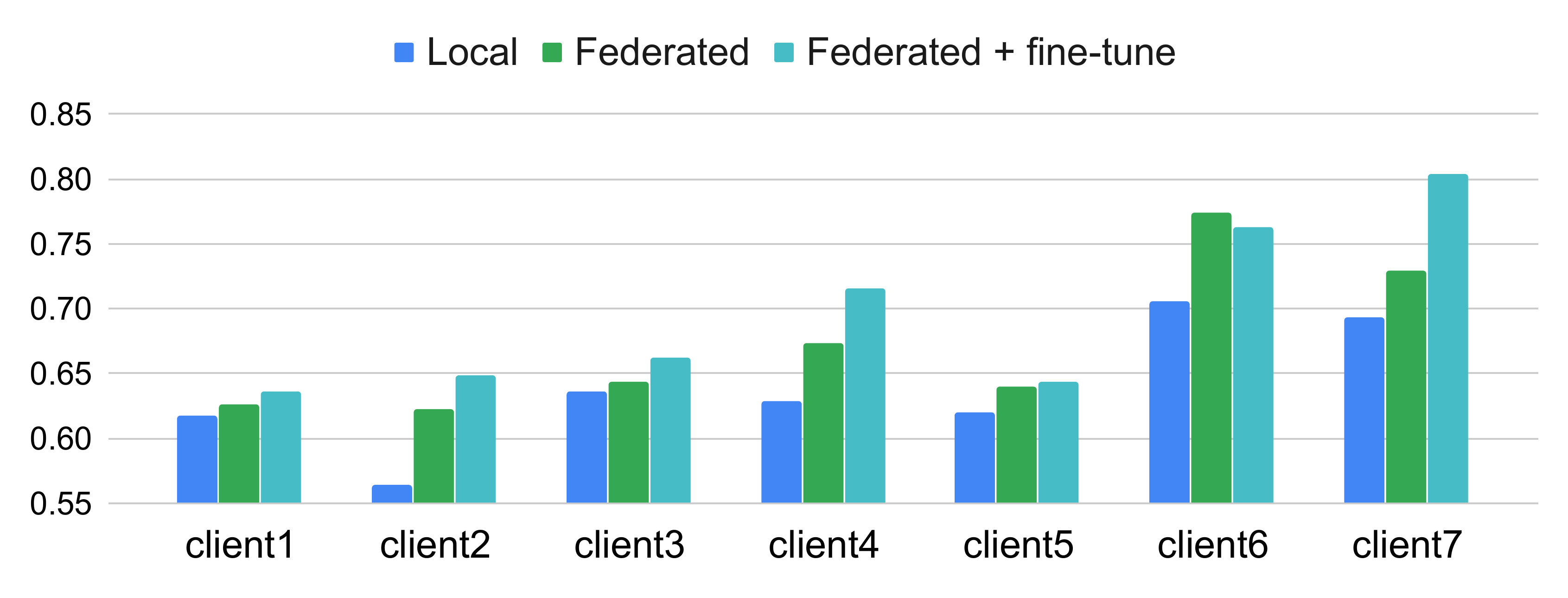} 
    \caption{Weighted linear kappa performance before and after federated learning, and after an additional round of local fine-tuning at each local site.}%
    \label{fig:mammography_results_fine-tuning}
\end{figure}

Fig. \ref{fig:mammography_results_fine-tuning} summarizes the kappa scores for local training, after FL, including after local fine-tuning, which improves a given model's performance on the client's local test data in all but one client.

\section{Discussion \& Conclusions}
\label{sec:conclusions}
Given our experimental results, we can see that federated learning (FL) in a real-world scenario can both achieve more accurate models locally as well as increase the generalizability of these models to data from other sources, such as test data from other clients. This improvement is due to the effectively larger training set made available through FL without the need to share any data directly.
While we cannot directly compare to a centralized training setting due to the nature of performing FL in a real-world setting, we observed that the average performance of models is similar to values reported in the literature on centralized datasets. For example, Lehman et al. \cite{lehman2019mammographic} reported a linear kappa value of 0.67 when applying DL for mammography breast density classification. We achieved an average performance of local models of 0.68 in the FL setting, confirming the ability of FL to achieve models comparable to models trained when the data is accumulated in a central database.
However, while the generalizability is improved, it is still not comparable to the performance on local test sets. In particular, the final global model is not near any acceptable performance on any of the local test datasets. The heterogeneity in results across institutions illustrates the difficulties in training models that are generalizable. In practice, some local adaptation (fine-tuning, see Fig. \ref{fig:mammography_results_fine-tuning}) or at least model selection based on local validation data (see diagonal of Table \ref{tab:mammography_generalizibility}) is needed. 

In this work, we deliberately did not attempt any data harmonization methods to study the effect of different data domains. The marked differences in intensity distributions due to different mammography systems are observable in Fig. \ref{fig:intensity_distribution}. Future work might explore the use of histogram equalization and other techniques \cite{karani2018lifelong,baweja2018towards} to harmonize non-IID data across different sites or investigate built-in strategies for domain adaptation within the FL framework \cite{li2020multi}. Similarly, we did not fully address issues of data size heterogeneity and class imbalance within our FL framework. For example, client 7 had almost no category (b) samples due to their local labeling practices required by their clinical protocol. Future work could incorporate training strategies such as client-specific local training iterations, other mini-batch sampling strategies, and loss functions.
We also did not attempt privacy-preservation techniques that would reduce the chance of model inversion and potential data leakage based on the trained models. Differential privacy could easily be applied to our framework, and it has been shown that it can achieve comparable results to the vanilla FL setting \cite{li2019privacy}.

Despite these challenges, we were able to train mammography models in a real-world FL setting that improved the performance of locally trained models alone, illustrating the promise of FL for building clinically-applicable models and sidestepping the need for accumulating a centralized dataset.


\paragraph{\textbf{Acknowledgements:}}
{
\footnotesize
Research reported in this publication was supported by a training grant from the National Institute of Biomedical Imaging and Bioengineering (NIBIB) of the National Institutes of Health (NIH) under award number 5T32EB1680 to K. Chang and J. B. Patel and by the National Cancer Institute (NCI) of the NIH under Award Number F30CA239407 to K. Chang. This study was supported by NIH grants U01CA154601, U24CA180927, U24CA180918, and U01CA242879, and National Science Foundation (NSF) grant NSF1622542 to J. Kalpathy-Cramer.
}
%
%
%

\clearpage
\newpage
\bibliographystyle{splncs04}
\bibliography{bibliography}

\begin{thebibliography}{10}
\providecommand{\url}[1]{\texttt{#1}}
\providecommand{\urlprefix}{URL }
\providecommand{\doi}[1]{https://doi.org/#1}

\bibitem{baweja2018towards}
Baweja, C., Glocker, B., Kamnitsas, K.: Towards continual learning in medical
  imaging. arXiv preprint arXiv:1811.02496  (2018)

\bibitem{boyd1995quantitative}
Boyd, N., Byng, J., Jong, R., Fishell, E., Little, L., Miller, A., Lockwood,
  G., Tritchler, D., Yaffe, M.J.: Quantitative classification of mammographic
  densities and breast cancer risk: results from the canadian national breast
  screening study. JNCI: Journal of the National Cancer Institute
  \textbf{87}(9),  670--675 (1995)

\bibitem{boyd2007mammographic}
Boyd, N.F., Guo, H., Martin, L.J., Sun, L., Stone, J., Fishell, E., Jong, R.A.,
  Hislop, G., Chiarelli, A., Minkin, S., et~al.: Mammographic density and the
  risk and detection of breast cancer. New England journal of medicine
  \textbf{356}(3),  227--236 (2007)

\bibitem{chang2018distributed}
Chang, K., Balachandar, N., Lam, C., Yi, D., Brown, J., Beers, A., Rosen, B.,
  Rubin, D.L., Kalpathy-Cramer, J.: Distributed deep learning networks among
  institutions for medical imaging. Journal of the American Medical Informatics
  Association  \textbf{25}(8),  945--954 (2018)

\bibitem{chang2020multi}
Chang, K., Beers, A.L., Brink, L., Patel, J.B., Singh, P., Arun, N.T., Hoebel,
  K.V., Gaw, N., Shah, M., Pisano, E.D., et~al.: Multi-institutional assessment
  and crowdsourcing evaluation of deep learning for automated classification of
  breast density. Journal of the American College of Radiology  (2020)

\bibitem{dunnmon2019assessment}
Dunnmon, J.A., Yi, D., Langlotz, C.P., R{\'e}, C., Rubin, D.L., Lungren, M.P.:
  Assessment of convolutional neural networks for automated classification of
  chest radiographs. Radiology  \textbf{290}(2),  537--544 (2019)

\bibitem{ho2014dense}
Ho, J.M., Jafferjee, N., Covarrubias, G.M., Ghesani, M., Handler, B.: Dense
  breasts: a review of reporting legislation and available supplemental
  screening options. American Journal of Roentgenology  \textbf{203}(2),
  449--456 (2014)

\bibitem{huang2017densely}
Huang, G., Liu, Z., Van Der~Maaten, L., Weinberger, K.Q.: Densely connected
  convolutional networks. In: Proceedings of the IEEE conference on computer
  vision and pattern recognition. pp. 4700--4708 (2017)

\bibitem{kaissis2020secure}
Kaissis, G.A., Makowski, M.R., R{\"u}ckert, D., Braren, R.F.: Secure,
  privacy-preserving and federated machine learning in medical imaging. Nature
  Machine Intelligence pp.~1--7 (2020)

\bibitem{karani2018lifelong}
Karani, N., Chaitanya, K., Baumgartner, C., Konukoglu, E.: A lifelong learning
  approach to brain mr segmentation across scanners and protocols. In:
  International Conference on Medical Image Computing and Computer-Assisted
  Intervention. pp. 476--484. Springer (2018)

\bibitem{landis1977measurement}
Landis, J.R., Koch, G.G.: The measurement of observer agreement for categorical
  data. biometrics pp. 159--174 (1977)

\bibitem{larson2020ethics}
Larson, D.B., Magnus, D.C., Lungren, M.P., Shah, N.H., Langlotz, C.P.: Ethics
  of using and sharing clinical imaging data for artificial intelligence: a
  proposed framework. Radiology p. 192536 (2020)

\bibitem{lehman2019mammographic}
Lehman, C.D., Yala, A., Schuster, T., Dontchos, B., Bahl, M., Swanson, K.,
  Barzilay, R.: Mammographic breast density assessment using deep learning:
  clinical implementation. Radiology  \textbf{290}(1),  52--58 (2019)

\bibitem{li2019privacy}
Li, W., Milletar{\`\i}, F., Xu, D., Rieke, N., Hancox, J., Zhu, W., Baust, M.,
  Cheng, Y., Ourselin, S., Cardoso, M.J., et~al.: Privacy-preserving federated
  brain tumour segmentation. In: International Workshop on Machine Learning in
  Medical Imaging. pp. 133--141. Springer (2019)

\bibitem{li2020multi}
Li, X., Gu, Y., Dvornek, N., Staib, L., Ventola, P., Duncan, J.S.: Multi-site
  fmri analysis using privacy-preserving federated learning and domain
  adaptation: Abide results. arXiv preprint arXiv:2001.05647  (2020)

\bibitem{mcmahan2017communication}
McMahan, B., Moore, E., Ramage, D., Hampson, S., y~Arcas, B.A.:
  Communication-efficient learning of deep networks from decentralized data.
  In: Artificial Intelligence and Statistics. pp. 1273--1282 (2017)

\bibitem{ooms2007mammography}
Ooms, E., Zonderland, H., Eijkemans, M., Kriege, M., Delavary, B.M., Burger,
  C., Ansink, A.: Mammography: interobserver variability in breast density
  assessment. The Breast  \textbf{16}(6),  568--576 (2007)

\bibitem{pisano2005diagnostic}
Pisano, E.D., Gatsonis, C., Hendrick, E., Yaffe, M., Baum, J.K., Acharyya, S.,
  Conant, E.F., Fajardo, L.L., Bassett, L., D'Orsi, C., et~al.: Diagnostic
  performance of digital versus film mammography for breast-cancer screening.
  New England Journal of Medicine  \textbf{353}(17),  1773--1783 (2005)

\bibitem{razzaghi2012mammographic}
Razzaghi, H., Troester, M.A., Gierach, G.L., Olshan, A.F., Yankaskas, B.C.,
  Millikan, R.C.: Mammographic density and breast cancer risk in white and
  african american women. Breast cancer research and treatment
  \textbf{135}(2),  571--580 (2012)

\bibitem{rieke2020future}
Rieke, N., Hancox, J., Li, W., Milletari, F., Roth, H., Albarqouni, S., Bakas,
  S., Galtier, M.N., Landman, B., Maier-Hein, K., et~al.: The future of digital
  health with federated learning. arXiv preprint arXiv:2003.08119  (2020)

\bibitem{sheller2018multi}
Sheller, M.J., Reina, G.A., Edwards, B., Martin, J., Bakas, S.:
  Multi-institutional deep learning modeling without sharing patient data: A
  feasibility study on brain tumor segmentation. In: International MICCAI
  Brainlesion Workshop. pp. 92--104. Springer (2018)

\bibitem{sickles2013acr}
Sickles, E., d’Orsi, C., Bassett, L., Appleton, C., Berg, W., Burnside, E.,
  et~al.: {ACR BI-RADS}{\textregistered} mammography. {ACR
  BI-RADS}{\textregistered} atlas, breast imaging reporting and data system
  \textbf{5}, ~2013 (2013)

\bibitem{spayne2012reproducibility}
Spayne, M.C., Gard, C.C., Skelly, J., Miglioretti, D.L., Vacek, P.M., Geller,
  B.M.: Reproducibility of bi-rads breast density measures among community
  radiologists: a prospective cohort study. The breast journal  \textbf{18}(4),
   326--333 (2012)

\bibitem{sprague2016variation}
Sprague, B.L., Conant, E.F., Onega, T., Garcia, M.P., Beaber, E.F., Herschorn,
  S.D., Lehman, C.D., Tosteson, A.N., Lacson, R., Schnall, M.D., et~al.:
  Variation in mammographic breast density assessments among radiologists in
  clinical practice: a multicenter observational study. Annals of internal
  medicine  \textbf{165}(7),  457--464 (2016)

\bibitem{sprague2014prevalence}
Sprague, B.L., Gangnon, R.E., Burt, V., Trentham-Dietz, A., Hampton, J.M.,
  Wellman, R.D., Kerlikowske, K., Miglioretti, D.L.: Prevalence of
  mammographically dense breasts in the united states. JNCI: Journal of the
  National Cancer Institute  \textbf{106}(10) (2014)

\bibitem{yaghjyan2011mammographic}
Yaghjyan, L., Colditz, G.A., Collins, L.C., Schnitt, S.J., Rosner, B., Vachon,
  C., Tamimi, R.M.: Mammographic breast density and subsequent risk of breast
  cancer in postmenopausal women according to tumor characteristics. Journal of
  the National Cancer Institute  \textbf{103}(15),  1179--1189 (2011)

\end{thebibliography}

\end{document}